
\input phyzzx

\def\d{\partial}
\def\r{\rho}
\def\rd{{\dot \rho}}
\def\rt{{\tilde \rho}}
\def\v{\upsilon}
\def\rv{{\rm v}}
\def\l{\lambda}
\def\rs{\rho_{\rm sol}}
\def\rw{\rho_{\rm wave}}

\Pubnum={CERN-TH.7496/94\cr}
\date={November 4, 1994}
\titlepage
\title{Waves and Solitons in the Continuum Limit of the
Calogero-Sutherland Model}
\bigskip
\author {Alexios P. Polychronakos\footnote\dagger
{poly@calypso.teorfys.uu.se}}
\address{Theory Division, CERN\break
CH-1211, Geneva 23, Switzerland\footnote*{Present address:
Theoretical Physics Dept., Uppsala University, S-751 08 Uppsala,
Sweden}}
\bigskip
\abstract{
We examine a collection of particles interacting with
inverse-square two-body potentials in the thermodynamic limit.
We find explicit large-amplitude density
waves and soliton solutions for the motion of the system. Waves
can be constructed as coherent states of either solitons or
phonons. Therefore, either solitons or
phonons can be considered as the fundamental excitations.
The generic wave is shown to correspond to a two-band state in
the quantum description of the system, while the limiting cases of
solitons and phonons correspond to particle and hole excitations.
}
\vskip 0.4in
\line {CERN-TH.7496/94\hfill}
\line {November 4, 1994\hfill}

\vfill
\endpage

\def\NP{{\it Nucl. Phys.\ }}

\def\PR{{\it Phys. Rev. \ }}
\def\PRL{{\it Phys. Rev. Lett.\ }}

\def\JP{{\it J. Phys.\ }}
\def\IJMP{{\it Int. Jour. Mod. Phys.\ }}

\REF\Calog{F.~Calogero, {\it Jour. of Math. Phys.}
{\bf10}, 2191 and 2197 (1969); {\bf 12}, 419 (1971).}
\REF\Suth{B.~Sutherland, \PR {\bf A4}, 2019 (1971); \PR {\bf A5}, 1372 (1972);
\PRL {\bf 34}, 1083 (1975).}
\REF\Mos{J.~Moser, ``Dynamical Systems, Theory and Applications,"
Lecture Notes in Physics {\bf 38}, Springer-Verlag, New York 1975.}
\REF\HalShas{F.~D.~M.~Haldane, \PRL {\bf 60}, 635 (1988);
B.~S.~Shastry,\PRL {\bf 60}, 639 (1988).}
\REF\CLP{H.~H.~Chen, Y.~C.~Lee and N.~R.~Pereira, {\it Phys.~Fluids}
{\bf 22} (1979) 187.}
\REF\SLA{ B.\ D.\ Simons, P.\ A.\ Lee, and B.\ L.\ Altshuler,
\PRL {\bf 70}, 4122 (1993);\PRL {\bf 72}, 64 (1994);
E.~Mucciolo, B.~Shastry, B.~Simons and B.~Altshuler,
cond-mat/9309030.}
\REF\FStat{ J.M.~Leinaas and J.~Myrheim, \PR {\bf B37}, 9286 (1988);
A. Polychronakos, \NP {\bf B234} (1989) 597;
F.D.M. Haldane, \PRL {\bf 67} (1991) 937;
Y.-S. Wu, \PRL {\bf 73} (1994) 922; S.~B.~Isakov, \IJMP {\bf A9} (1994) 2563.}
\REF\SC{B.~Sutherland and J.~Campbell (preprint), Sept. 1993.}
\REF\BZ{N.~N.~Bogoliubov and D.~N.~Zubarev, JETP 1 (1955) 83;
A.~Jevicki and B.~Sakita, \NP {\bf B165} (1980) 511.}
\REF\AJL{I.~Andric, A.~Jevicki and  H.~Levine, \NP {\bf B215} (1983) 307;
I.~Andric and V.~Bardek, \JP {\bf A21} (1988) 2847.}
\REF\J{A.~Jevicki, \NP {\bf B376} (1992) 75.}
\REF\Pol{J.~Polchinski, \NP {\bf B362} (1981) 125.}
\REF\MP{J.~Minahan and A.~P.~Polychronakos, \PR {\bf B50} (1994) 4236.}

{\it 1.~Introduction and basic setup:} There has been much recent
interest in the Calogero-Moser-Sutherland model of interacting particles
in one dimension [\Calog,\Suth,\Mos] (which is often referred to in
the physics literure as the CS model).
This model is related to quantum spin chains with long range
interactions between the spins [\HalShas], wave propagation in
stratified fluids [\CLP], random matrix theory [\Suth,\SLA] and
fractional statistics [\FStat].

The CS model is exactly solvable in
both the classical and the quantum regime. Remarkably, the quantum
solution is much easier to interpret, exhibiting a straightforward
analogy to the free fermion case. In a recent paper, Sutherland and
Campbell examined the classical system in the thermodynamic limit
and identified the excitations [\SC]. It was found that the classical
system has solitons, corresponding to a single particle running through
the rest of them, as well as small amplitude waves (phonons), identified
with holes. The purpose of this paper is to derive large amplitude wave
and soliton solutions of the classical system in the continuous limit,
where the particles form a ``fluid," and examine their correspondence to
the quantum states.

We consider a collection of particles with the hamiltonian
$$
H = \half \sum_{i=1}^N {\dot x}_i^2 + \sum_{i>j} {g \over ( x_i - x_j )^2}
\eqn\H$$
where for convenience we chose them of unit mass. In principle, such a
system can be put in a box of length $L$ (with an appropriate modification
of the potential [\Suth]). We shall be interested in the limit $N,L \to \infty$
with $N/L$ fixed. In this limit, the system can be described in terms of
a density field $\r (x)$ and a velocity field $\v (x)$.
At equilibrium, the particles will form
a regular lattice of spacing $a$ and density $\r_o = 1/a$. The particle
current is $J=\r \v$ and by particle conservation
$$
\rd + \d J = \rd + \d (\r \v)=0
\eqn\cons$$
where $\d = \d / \d x$. The kinetic energy of the system is
$$
K=\int dx \half \r \v^2
$$
We can formally solve \cons\ for $\v$ to obtain $\v = -\d^{-1} \rd / \r$,
and the expression for the kinetic energy becomes
$$
K = \int dx {(\d^{-1} \rd )^2 \over 2\r}
\eqn\kin$$
This is exactly the kinetic term of the collective field hamiltonian
description of a many-body system [\BZ]. The potential energy can also be
expressed in terms of the density. The naive expression, however, which
would be
$$
V= \int dx dy ~{g\over 2} {\r(x) \r(y) \over (x-y)^2}
$$
is incorrect. The reason is that the interaction is singular at
coincidence points, and thus a substantial part of the potential
energy comes from nearest neighbors and is not accurately
reproduced by the naive continuous expression. The correct expression
requires a careful conversion of the discrete sum in terms of the
continuous fields. Alternatively, we can simply take the classical
limit ($\hbar \to 0$) of the quantum mechanical expression derived
in the collective field formulation [\AJL]. The result is
$$
V = \int dx \left\{ {\pi^2 g \over 6} \r^3 - {g\over 2} \r \d \rt
+ {g\over 8} {(\d\r)^2 \over \r} \right\}
\eqn\pot$$
where $\rt$ stands for the Hilbert transform:
$$
\rt = \int dy ~P.P.{1\over x-y} ~\r(y) = \half \lim_{\epsilon \to 0}
\int dy ~({1\over x-y+i\epsilon} + {1\over x-y-i\epsilon}) ~\r(y)
\eqn\H$$
The first term, which accounts for the interaction of each particle with
its few nearest neighbors, is the dominant one in the limit where the
scale of variation of $\r$ is much larger than the lattice spacing.
In our case, however, we are interested in finite-width fluctuations
and we must keep the full expression.

The dynamics of the system can be found by varying the lagrangian
$L=K-V +\mu \r$ with respect to $\r$. The chemical potential $\mu$
plays the role of a Lagrange multiplier ensuring that the total number of
particles remains constant. The resulting equations of motion are
$$
-\d^{-1}{\dot \v} - \half \v^2 -{\pi^2 g\over 2} \r^2 +g~\d\rt
+{g\over 8} \Bigl( {\d\r \over \r}\Bigr)^2 + {g\over 4} \d \Bigl(
{\d\r \over \r} \Bigr) + \mu = 0
\eqn\em$$
as well as \cons. The inverse derivative operator in \em\ is defined
in terms of the principal value, in Fourier space $\d^{-1} = \lim_{\epsilon
\to 0} k/(k^2 + \epsilon^2)$. In particular, acting on a constant it gives
zero. By requiring that the static configuration $\v=0$, $\r = \r_o$ be a
solution of \em, we obtain the value of the chemical potential
$$
\mu = {\pi^2 g \over 2} \r_o^2
\eqn\chem$$
This is in agreement with the value obtained from the exact solution
of the many-body problem [\Suth,\SC].

{\it 2.~Small-amplitude waves:} From the above equations we can obtain
the dispersion relation in the linearized regime of small-amplitude waves,
which we shall call phonons.
Noting that the Fourier transform of $\d\rt$ is $\pi |k| \r(k)$, we obtain
$$
\rv_{{\rm phase}}^2 = \Bigl({\omega \over k} \Bigr)^2 =
g \left(\pi \r_o - {|k| \over 2}\right)^2 ~~~~{\rm or}~~~~
\omega = \sqrt{g}\left(\pi \r_o |k| - {k^2 \over 2}
\right)
\eqn\disp$$
{}From \disp\ we deduce that the velocity of sound $\rv_s$, defined as the
phase (or group) velocity in the long wavelength limit, is
$$
\rv_s = \pi \r_o \sqrt{g}
\eqn\vs$$
In terms of the group velocity $\rv_g$ the dispersion relation becomes
$$
\omega = {\rv_s^2 - \rv_g^2 \over 2\sqrt{g}}
\eqn\group$$
We observe that \vs\ and \group\ are the exact results. The group
velocity is always smaller that the velocity of sound, and the above
linearized waves can be identified with holes in the quantum theory.
Notice that the above formulae are valid for $|k| < \pi \r_o =
{\pi \over a}$, else the group velocity turns negative. This is reasonable,
since the above condition restricts the momentum to
the fundamental region of the Brillouin zone, thus avoiding umklapp.

{\it 3.~Solitons:} As observed in [\SC], the many-body system should
exhibit soliton solutions, corresponding to particle excitations.
On the other hand, in [\J] an equation similar to \em\ was written
for a system of free fermions, coming from an effective lagrangian
chosen so as to reproduce the full quantum mechanical dispersion
relation of the system at the semiclassical level. This equation has
solitary wave solutions [\J]. As we will demonstrate here, our equations
\em, \cons\ also have solitary wave solutions of a rational type;
we shall call these solutions solitons, and will comment later on
their true nature.

For a localized constant profile configuration, propagating at speed
$\rv$, both $\r$ and $\v$ are functions of $x-\rv t$ only.
{}From \cons\ we have
$$
\d(\v \r - \rv \r ) = 0 ~~~~{\rm and~thus}~~~~ \v = {\r - \r_o \over
\r} \rv
\eqn\vv$$
In the above, the integration constant is fixed by the boundary
condition that $\v \to 0$ at $x \to \pm \infty$, where $\r \to \r_o$.
Similarly, \em\ becomes
$$
{\rv^2 \over 2} \Bigl({\r_o^2 \over \r^2} - 1 \Bigr) + {\pi^2 g \over
2} (\r^2 - \r_o^2 ) - g~\d \rt -{g \over 8} \Bigl({\d\r \over \r}
\Bigr)^2 - {g \over 4}\d \Bigl({\d\r \over \r}\Bigr) = 0
\eqn\soleq$$
To guess a solution for \soleq\ of the form $\rs = \r_o + \delta
\r$, where $\delta \r$ is localized, we notice that the term in
\soleq\ containing the Hilbert transform will always produce out
of a localized function a tail falling off quadratically.
Thus, $\delta\r$ itself should have such a behavior at infinity.
The simplest function of this form is
$$
\rs = \r_o + {A \over x^2 + B^2}
$$
Plugging the above form in \soleq\ we find, after an amount of
algebra, that it is indeed a solution, provided that $\rv > \rv_s$
and
$$
A = {u \over \pi^2 \r_o}~,~~ B = {u \over \pi \r_o} ~~~~
{\rm where}~~~~ u = {\rv_s^2 \over \rv^2 - \rv_s^2}
\eqn\ABu$$
We finally arrive at the soliton profile
$$
\rs = \r_o \left( 1 + {u \over (\pi \r_o x)^2 + u^2} \right)~,
{}~~~~ u = {\rv_s^2 \over \rv^2 - \rv_s^2}
\eqn\solsol$$
The above solution is, strictly speaking, a solitary wave.
True solitons are solitary wave solutions of integrable
equations, and scatter off each other preserving their number
and asymptotic momenta. Since the initian many-body system (1)
is integrable, we expect the corresponding continuum system
to be also integrable, although a direct prrof is lacking,
and thus \solsol\ to be a true soliton. This is corroborated
by the correspondence of these solutions to particles, as
demonstrated below.

The above soliton carries particle number $Q$, momentum $P$
and energy $E$, defined as the extra amount over the static
solution $\r_o$. We find
$$
\eqalign{
Q &= \int dx ~(\rs - \r_o) = 1 \cr
P &= \int dx ~\rs ~\v = \rv \cr
E &= \int dx ~[K(\rs)+V(\rs) - V(\r_o)] = \half \rv^2}
\eqn\QPE$$
We observe that the net particle number carried by the soliton
is 1, independently of its velocity; its momentum and energy are
also those of a free particle of unit mass moving at the soliton
velocity $\rv$. Therefore, the soliton can be exactly identified
with a particle excitation of the system. Again, this is in agreement
with exact results drawn from the quantum theory, where particle
excitations always move faster than the sound [\SC]. Notice, further,
that the solitons become thinner as their velocity increases, while
their spread diverges as they slow down to the velocity of sound.

The above result for $Q$ implies that the displacement of the
equilibrium lattice far away from the soliton is $\pm$ half lattice
spacing either way (so that there is an excess of one particle near
the soliton). This result, as well as the form of the soliton \solsol,
are at odds with the results found in [\SC]. We suspect that the source of
the discrepancy is the truncation to a finite number of $x$-derivatives
of the form for the potential in [\SC]; this turns the equation to a
local one and gives the soliton an exponential decay, rather than the
inverse-square decay of the nonlocal equation. We also notice that our
soliton has some important qualitative differences from the solitons
in the semiclassical fermion theory of [\J]: Our solitons carry a positive
particle number of 1, as opposed to a negative particle number in [\J],
which would rather identify them as holes. Further, there are no static
solitons in our case, since $|\rv| > \rv_s$, while in [\J] solitons
can slow down to zero speed. Finally, the definition of momentum used
in [\J] differs from ours by a surface term. Clearly \QPE\ is the
physically sensible definition in our case.

{\it 4.~Finite amplitude waves:} Soliton profiles moving at very large
distances from each other will obviously remain solutions. If we
could form a state consisting of a sequence of solitons at regular
distances spaced by $\l$, all moving with the same velocity $\rv$, we
would have found a large-amplitude wave solution with wavelength $\l$.
We thus try the form
$$
\rw - \r_o = \sum_{n=-\infty}^\infty \Bigl( \rs (x-n\l) -\r_o \Bigr) =
{1\over \l} ~{\sinh{2u \over \l \r_o} \over \cosh{2u \over \l \r_o} -
\cos{2\pi x \over \l}}
\eqn\w$$
where now the parameter $u$ is {\it not} necessarily given by $\rv_s^2
/ (\rv^2 - \rv_s^2)$, since the proximity of the other solitons may have
changed their common velocity. The above waveform is characterized by its
amplitude $A$, defined as midway the distance from peak to trough,
$$
A= {\r_{\rm max} - \r_{\rm min} \over 2} = {1 \over \l \sinh{2u
\over \l \r_o}}
\eqn\A$$
as well as by its wavelength $\l$.
Substituting the form \w\ in \soleq\ we find, again after quite a bit of
algebra, that it is indeed a solution provided
$$
\tanh{2u \over \l \r_o} = {2 \l \r_o \rv_s^2 \over \l^2 \r_o^2 (\rv^2 -
\rv_s^2 ) - \rv_s^2 }
\eqn\tann$$
The above is the amplitude-dependent dispersion relation for the nonlinear
waves of the system. Before we interpret it, however, we must note the
following: The conventions used for deriving \soleq \ were that the
solution $\r$ carries some particle number and momentum on top of the
``vacuum" solution $\r_o$. This is reasonable for an isolated soliton,
but rather inconvenient for a wave solution, which is thought to
be a fluctuation carrying no net particle number and no net momentum
(no drift). But the presence of the solitons in \w\ adds one particle
per length $\l$, and thus the true equilibrium density of the system
is $\r_o + {1\over \l}$.
Further, the solitons contribute a momentum $\rv$ per
length $\l$; to neutralize it, we must boost the whole system in the
opposite direction by an appropriate amount. After performing these
redefinitions, the expression for the wave in terms of the true
velocity $\rv$ and true background density $\r_o$ is
$$
\rw = \r_o + {1\over \l} \left( {1 \over \sqrt{\l^2 A^2 + 1} -
\l A \cos{2\pi x \over \l}} - 1 \right)
\eqn\wsol$$
and the nonlinear dispersion relation in terms of the amplitude $A$ is
$$
\rv = {\omega \over k} = \Bigl( \rv_s - {\pi \sqrt{g} \over \l} \Bigr)
\sqrt{1+{2 A^2 (\l \r_o - 1) \over \r_o^2 (1 + \sqrt{\l^2 A^2 +1} )}}
\eqn\Adisp$$

In the limit $\l \to \infty$ the above equations reduce to the single
soliton solution. In the limit $A \to 0$, on the other hand, the above
formulae become
$$
\eqalign{
\rw &= \r_o + A \cos{kx} ~,~~~~~ k={2\pi\over \l} \cr
\rv &= {\omega \over k} = \rv_s - {\sqrt{g} \over 2} k}
\eqn\Asmall$$
which is the small amplitude wave solution and dispersion relation.
We see, therefore, that the above solutions interpolate between the
two extreme cases. We stress that the generic wave can run either
faster or slower than the speed of sound.
It should also be noted that the above wave solution constitutes a
solitary wave for the continuum limit of the system with periodic space
(that is, the Sutherland model), where the period is the wavelength.

{\it 5.~Discussion and conclusions:}
In summary, we have found exact soliton and wave solutions for the
CS system in the continuum limit. Certainly the above do
not exhaust the list of solutions; the general motion of the system
will be a nonlinear superposition of waves (or solitons). Although
we could find such many-soliton or many-wave solutions, it is an
algebraically laborious task of not much interest. It serves,
nevertheless, as an indication
that the above solitary waves are true solitons, as expected from the
integrability of the original model.

It is instructive to put the above solutions into correspondence
with the quantum mechanical states. Consider $N$ particles in a space
of length $L$. The ground state of the system consists of a ``Luttinger
sea" in the pseudomomentum, with spacing between adjacent particles equal
to $2\pi \ell /L$ and ``Fermi level" $\pi \ell N/L$, where
$g=\ell (\ell -\hbar )$.
At the limit $\hbar \to 0$, $N,L \to \infty$, $N/L \to \r_o$, the
ground state becomes a continuous filled band with Fermi level $P_F =
\pi \sqrt{g} \r_o$.  A small amplitude wave, corresponding to a hole,
is a very small gap in the band. A soliton, corresponding to a particle
excitation, is a single particle peeled from the Fermi level and
placed some distance above. The generic finite amplitude wave corresponds
to a state with {\it two} continuous filled bands, of widths $P_1$ and
$P_2$ (with $P_1 + P_2 = 2 P_F$) and with a gap $G$ between them. These
are related to the wave parameters as
$$
\eqalign{
\l &= {2\pi \sqrt{g} \over P_1 } \cr
\rv &= {P_2 \over 2} \left({G \over \pi \sqrt{g} \r_o} + 1\right)}
\eqn\qu$$
Such a state can be visualized as arising either by successively
exciting single particles by the same constant momentum, until they
form a continuous band, or by gradually augmenting the gap of a hole,
until it becomes finite. This state can thus be thought of as either
a coherent state of solitons (much like the way we constructed the
wave solution), or as a coherent state of  phonons, their nonlinear
nature accounting for the change in profile as they accumulate.
Indeed, the soliton itself can be thought of as a superposition of
many phonons with very large wavenumber, and the phonon as a superposition
of many solitons just above the Fermi level. For the finite $N$ (finite
$L$) system the distinction between the two is fuzzy and in principle
only one kind of excitations need be considered as fundamental. Note,
further, that quantum mechanically the holes behave as particles with
fractional statistics of order $\hbar /\ell$ (meaning that $\ell/\hbar$
of them put together would form a fermion). At the classical limit
$\hbar \to 0$, thus, they become bosons, as they should be since
phonons obey no exclusion principle. Particles, on the other hand,
carry statistics of order $\ell/\hbar$. Thus in the classical limit
they become ``superfermions" meaning that no two of them can occupy
relatively nearby quantum states. This is consistent with the inverse
square repulsion between the classical particles.

The above results are of direct relevance to the large-$N$ limit
of one-dimensional free matrix models. The particular wave and
soliton solutions correspond to motions of the density of
eigenvalues in the unitary and hermitian models, respectively.
Taking, for clarity, the hermitian case, the motion of a free
$N \times N$ matrix $M$ with angular momentum $\ell$ is
$$
M_{jk} = \delta_{jk} (p_j t + a_j ) + (1-\delta_{jk} )
{i\ell \over p_j - p_k}
\eqn\M$$
The situation where most of the eigenvalues lie on a regular
lattice with only one of them moving with velocity v is reproduced
by choosing
$$
p_j = {2\pi \ell \over a(N-2)}\Big(j-{N\over 2}\Big) ~~
({\rm for}~j<N) ~,~~~~ p_N = \rv ~,~~~~~a_j =0
\eqn\pa$$
(Notice that the above momenta $p_1 , \dots p_{N-1}$ span the values
between the two ``Fermi'' levels $\pm {\pi \ell \over a}$.)
It should be possible to prove analytically that the eigenvalues
of \M\ with parameters \pa\ have a density as given by our
soliton solution, but in practice this is a very hard task.
The corresponding problem for unitary matrices is even harder
to tackle, while our wave \wsol\ readily provides the solution.
Many-soliton solutions will be given by eigenvalues of \M\
with, now, more than one of the momenta $p_j$ taking values
equal to the velocities of the solitons, while the rest span the
``Luttinger sea''.

The solutions found in this paper are very similar to the ones in
stratified fluids. This is sensible, since the motion of these
fluids (under some conditions) is governed by the Benjamin-Ono
equation, which is known to have solitons behaving like Calogero
particles [\CLP]. This also suggests that the the many-soliton
solutions of the CS model will correspond to the ones of the
Benjamin-Ono equation, at least when all of them are left-
or right-movers. The interesting fact is that stratified fluids
themselves behave, in this respect, as hydrodynamic collections of
CS particles. The exact mathematical connection of the two
systems is still obscure.

We conclude by noting that the quantum mechanical problem separates
into two noninteracting chiral sectors, having to do with excitations
near either end of the Luttinger sea. (The two sectors mix nonperturbatively
when a number of particles of order $N$ is excited, depleting the sea.)
Therefore, the equation \em\ governing the continuum system should also
decompose into two nonmixing, first-order in time equations, one for
each sector. For the corresponding equation for free fermions this is
indeed the case [\Pol]. In fact, from the collective field description of
the system when only one chiral sector is present, we deduce that the chiral
equations are exactly of the Benjamin-Ono type [\AJL,\MP].
The exact field combinations in terms of which
this decomposition would be achieved, however, are not known and
constitute an open problem.

\ack{I would like to thank Bill Sutherland and Joe Minahan
for discussions, and the Aspen Center of Physics for its
hospitality during summer 1994, where this work was started.}

\refout
\end